\documentclass[11pt,reqno,twoside]{article}

\usepackage[T1]{fontenc}

\usepackage[]{amsmath, amssymb, amsthm, tabularx}

\usepackage[]{a4, xcolor, here}

\usepackage[]{pstricks, pst-text, pst-node, pst-tree, gastex}

\usepackage{tikz}

\usepackage[tikz]{bclogo}
\usepackage{comment}

\usepackage{boites,graphicx}

\usepackage{soul}

\definecolor{pastelyellow}{rgb}{0.99, 0.99, 0.59}
\definecolor{aqua}{rgb}{0.0, 1.0, 1.0} 
\definecolor{aquamarine}{rgb}{0.5, 1.0, 0.83} 
\definecolor{bananayellow}{rgb}{1.0, 0.88, 0.21}
\definecolor{burgundy}{rgb}{0.5, 0.0, 0.13}
\definecolor{ao(english)}{rgb}{0.0, 0.5, 0.0}

\setul{}{0.2ex}
\setulcolor{bananayellow}

\usepackage[normalem]{ulem}

\usepackage{mathrsfs}

\usepackage{stmaryrd}

\usepackage{enumerate}

\usepackage{paralist}

\usepackage{multienum}

\usepackage{multicol}

\usepackage{fancyhdr}

\usepackage{datetime}

\usepackage{everypage}

\usepackage[contents={},opacity=1,scale=1.6,
color=gray!90]{background}

\usepackage{ifthen}

\usepackage[]{hyperref}

\hypersetup{
		colorlinks = true,
		linkcolor = red,
		anchorcolor = black,
		citecolor = blue, 
		filecolor = cyan,
		menucolor = red,
		runcolor = cyan,
		urlcolor = blue,
		linkbordercolor = {white},
		linktocpage = true
}

\newtheorem{Theorem}{Theorem}[section]
\newtheorem{Proposition}[Theorem]{Proposition}

\newtheorem{Corollary}[Theorem]{Corollary}

\theoremstyle{Definition}
\newtheorem{Definition}[Theorem]{Definition}

\newtheorem{Example}[Theorem]{Example}
\newtheorem{Remark}[Theorem]{Remark}

\makeatletter
\def\thmhead@plain#1#2#3{%
  \thmname{#1}\thmnumber{\@ifnotempty{#1}{ }\@upn{#2}}%
  \thmnote{ {\the\thm@notefont#3}}}
\let\thmhead\thmhead@plain
\makeatother

\newcommand{\cC}{\mathcal{C}}

\newcommand{\cE}{\mathcal{E}}
\newcommand{\cF}{\mathcal{F}}
\newcommand{\cG}{\mathcal{G}}

\newcommand{\cP}{\mathcal{P}}

\newcommand{\cU}{\mathcal{U}}
\newcommand{\cV}{\mathcal{V}}

\newcommand{\cX}{\mathcal{X}}

\newcommand{\type}{(t_1, \ldots, t_r)}

\newcommand{\bbF}{{\mathbb F}} 

\renewcommand{\geq}{\geqslant}
\renewcommand{\leq}{\leqslant}

{\end{list}}%

\begin{document}

\renewcommand{\headrulewidth}{0pt}

\rhead{ }
\chead{\scriptsize Consistent Flag Codes}
\lhead{ }

\title{Consistent Flag Codes}

\author{\renewcommand\thefootnote{\arabic{footnote}}
Clementa Alonso-Gonz\'alez\footnotemark[1],\,  Miguel \'Angel Navarro-P\'erez\footnotemark[1]}

\footnotetext[1]{Dpt.\ de Matem\`atiques, Universitat d'Alacant, Sant Vicent del Raspeig, Ap.\ Correus 99, E-03080 Alacant. \\ E-mail adresses: \texttt{clementa.alonso@ua.es, miguelangel.np@ua.es}.}

\date{}

\maketitle

\begin{abstract}

In this paper we study flag codes on $\bbF_q^n$, being $\bbF_q$ the finite field with $q$ elements. Special attention is given to the connection between the parameters and properties of a flag code and the ones of a family of constant dimension codes naturally associated to it (the \emph{projected codes}). More precisely, we focus on \emph{consistent flag codes}, that is, flag codes whose distance and size are completely determined by their projected codes. We explore some aspects of this family of codes and present examples of them by generalizing the concepts of \emph{equidistant} and \emph{sunflower} subspace code to the flag codes setting. Finally, we present a decoding algorithm for consistent flag codes that fully exploits the consistency condition.

\end{abstract}



\section{Introduction}
The concept of {\em network coding} was introduced in \cite{AhlsCai00} as a method to increase the information flow within a network modelled as an acyclic directed graph with possibly more than one source and receiver. This network operates with vectors of a given vector space $\bbF_q^n$ over the finite field of $q$ elements $\bbF_q$, being $q$ a prime power. The intermediate nodes transmit random  $\mathbb{F}_q$-linear combinations of these vectors, instead of simply routing them. In \cite{KoetKschi08},  Koetter and Kschischang presented an algebraic approach to network coding. Since vector spaces are invariant by linear combinations, the authors suggested using vector subspaces, in lieu of vectors, as codewords. In the same paper, the authors explained how to use the channel in order to send  a vector space of $\mathbb{F}_q^n$. The sender injects the set of vectors of any basis of the given vector space into the network and every intermediate node sends random $\mathbb{F}_q$-linear combinations of the available vectors. In the end, the receiver collects the incoming vectors and forms the $\mathbb{F}_q$-vector subspace spanned by them. In this context, a \emph{subspace code} of length $n$ is just a nonempty collection of subspaces of $\bbF_q^n$. In case we restrict ourselves to subspaces with the same dimension, we speak about \emph{constant dimension codes}. The study of constant dimension codes has lead to many papers in recent years. We refer the reader to \cite{TrautRosen18} and references therein for the basics on these codes.

Subspace codes require a single use of the channel described above to send a codeword, i.e., a subspace. These codes were generalized in \cite{NobUcho09} as the so-called \emph{multishot} subspace codes. More precisely, in an \emph{$r$-shot} code, codewords are sequences of $r\geq 2$ vector subspaces of $\mathbb{F}_q^n$. In this case, sending a codeword needs $r$ uses (shots) of the channel. As it was shown in that paper, fixed the values $n$ and $q$, multishot codes could achieve better cardinality and distance than \emph{one-shot} codes just by introducing a new parameter: the number of the channel uses. 

In this paper we focus on \emph{flag codes}, a specific family of multishot codes whose codewords are given by sequences of nested subspaces (\emph{flags}) with prescribed dimensions. In the network coding framework, flag codes were introduced in \cite{LiebNebeVaz18}. In that work, the authors studied flag codes as orbits of subgroups of the general linear group and provided some constructions of them as well as a new channel model for flags.

The goal of the present work is the study of the connection between  the parameters and properties of a flag code and the ones of its \emph{projected codes}, that is, the constant dimension codes used at each shot when sending flags of a flag code. In this direction, we introduce the concept of \emph{consistent flag codes}, a family of flag codes whose cardinality and distance are perfectly described in terms of its projected codes. This notion of \emph{consistency} (\emph{cardinality-consistency} together with \emph{distance-consistency}) will allow us to easily translate distance and cardinality properties of a flag code to the subspace code level and vice versa. Moreover, in a consistent flag code, some structural properties satisfied at flag codes level are transferred as the equivalent properties at the subspace codes level, that is, they are properly inherited by the projected codes (and conversely). We will exhibit this fact providing two specific families of consistent flag codes coming from the natural generalization of \emph{equidistant} and \emph{sunflower} constant dimension codes (see \cite{EtzRav15, GoRav16}). The consistency condition will be exploited to give a decoding algorithm, which translates the problem of decoding a flag code to the one of decoding a constant dimension code.  

The paper is is organized as follows. In Section \ref{sec:preliminares}, we provide the basic background on subspace codes, focusing on two well-known families of constant dimension codes: equidistant and sunflower codes. Besides, some definitions and known facts about flag codes are presented, together with the channel model to be used later on. Section \ref{sec:Consistent flag codes} is devoted to properly define the concept of consistency of a flag code with respect to its projected codes. In Section \ref{sec: Some families of consistent flag codes}, we present some families of consistent flag codes by generalizing the concepts of equidistant and sunflower code to the flag codes scenario. Furthermore, we will see that the only consistent equidistant (resp. sunflower) flag codes are the ones that have equidistant (resp. sunflower) projected codes. Finally, in Section \ref{sec: Our decoding algorithm}  we study the problem of decoding consistent flag codes on the erasure channel by exhibiting a suitable decoding algorithm.

\section{Preliminaries}\label{sec:preliminares}

This section is devoted to recall some background needed along this paper. The first part concerns subspace codes, focusing on two important families of constant dimension codes. In the second part we remind some known facts and definitions related to flag codes. 

\subsection{Subspace codes}\label{subsec: subspace and multishot codes}

Let $q$ be a prime power and $\bbF_q$ the finite field of $q$ elements. For every $n\geq 1$, we denote by $\cP_q(n)$ the \emph{projective geometry} of the vector space $\bbF_q^n$, which consists of the set of all the $\bbF_q$-vector subspaces of $\bbF_q^n$. This set can be endowed with a metric, the \emph{subspace distance}, given by
\begin{equation}\label{eq: distancia subespacios}
d_S(\cU, \cV)= \dim(\cU+\cV) - \dim(\cU\cap\cV), \ \forall \, \cU, \cV \in \cP_q(n).
\end{equation}
The \emph{Grassmannian} $\cG_q(k,n)$ (or Grassmann variety) of dimension $k\leq n$ of $\bbF_q^n$ is just the set of $k$-dimensional subspaces in $\cP_q(n)$. The subspace distance induces in turn a metric in $\cG_q(k,n)$ and, in this case, its expression  becomes
$$
d_S(\cU, \cV)=2(k - \dim(\cU\cap\cV)), \ \forall \, \cU, \cV \in \cG_q(k,n).
$$ 

A \emph{subspace code} of $\bbF_q^n$ is a nonempty subset $\cC$ of $\cP_q(n).$ If every subspace in $\cC$ has the same dimension, say $k$, then it is said to be a \emph{constant dimension code} in the Grassmannian $\cG_q(k, n)$. The \emph{(minimum) distance} of a subspace code $\cC$ is 
$$
d_S(\cC)= \min\{ d_S(\cU, \cV) \ | \ \cU, \cV\in \cC, \ \cU\neq\cV \}.
$$
For codes consisting of just one element, we put $d_S(\cC)=0$. In any other case, the minimum distance of a subspace code is a positive integer. For a constant dimension code $\cC$ in the Grassmannian $\cG_q(k, n)$, the minimum distance $d_S(\cC)$ is an even integer with
\begin{equation}\label{eq:bound subspace codes}
    d_S(\cC) \leq \min\{2k, 2(n-k)\}.
\end{equation} 
For the basic background on constant dimension codes we refer the reader to \cite{KoetKschi08}, the seminal paper in this subject, and to \cite{TrautRosen18}.

A constant dimension code $\cC \subseteq \cG_q(k,n)$ is said to be \emph{equidistant} if its distance is attained by every pair of different subspaces in $\cC$. In this situation, there exists an integer $c$ such that $\max\{0, 2k-n\}\leq c\leq k$ and $d_S(\cC)=2(k-c)$. This value $c$ represents the dimension of the intersection between every pair of different subspaces in $\cC$. Due to this reason, these codes are also known as \emph{equidistant $c$-intersecting} constant dimension codes. Trivial codes consisting just of one element are trivially equidistant of distance zero. Equidistant subspace codes have been widely studied in \cite{EtzRav15, GoRav16}. In Section \ref{sec:Consistent flag codes} we will generalize this concept to the flag codes setting.

Observe that constant dimension codes in the Grassmannian $\cG_q(k,n)$ attaining the maximum distance are, in particular, equidistant $c$-intersecting constant dimension codes with $c=\max\{0, 2k-n\}$. For dimensions up to $\lfloor \frac{n}{2}\rfloor$, they are better known as \emph{partial spread codes}. For further information on this family of codes, consult \cite{GoRav14}.

Regarding the intersections between couples of codewords, there is another interesting class of equidistant constant dimension codes to take into account. A constant dimension code $\cC\subseteq\cG_q(k,n)$ is said to be a \emph{sunflower} if there exists a subspace $C$ such that, for every pair of different subspaces $\cU, \cV\in \cC$, it holds $\cU\cap\cV=C$. In this case, the subspace $C$ is called the \textit{center of the sunflower}. Observe that a sunflower is an equidistant $c$-intersecting code with $c=\dim(C)$. These codes have been also studied and constructed in \cite{EtzRav15, GoRav16}. Observe that every subspace code with just one codeword $\cC=\{\cU\}$ can be seen as a trivial sunflower of center $\cU$.

Concerning the way we use the channel to transmit some information encoded in a subspace code, recall that, to send a codeword (a subspace) we just need to use the channel once, that is, we perform one \emph{shot} (see \cite{KoetKschi08}). Under this viewpoint, subspace codes can be called \emph{one-shot} codes. In contrast, when the number of required uses of the channel is bigger, say $r$, we speak about \emph{multishot codes} of length $r$ or, simply, \emph{$r$-shot} codes. More precisely, $r$-shot codes are nonempty subsets of $\cP_q(n)^r$, i.e., their codewords are sequences of length $r$ of subspaces of $\bbF_q^n$. The subspace distance defined in (\ref{eq: distancia subespacios}) can be naturally generalized to this setting. Given two sequences of subspaces $\cU=(\cU_1, \dots, \cU_r)$ and $\cV=(\cV_1, \dots, \cV_r)$, their \emph{extended subspace distance} is given by
\begin{equation}\label{eq: extended subspace distance}
d_S(\cU, \cV)= \sum_{i=1}^r d_S(\cU_i, \cV_i).
\end{equation}
For further information on multishot codes, see \cite{NobUcho09}.

\subsection{Flag codes}\label{subsec: flags and channel}

\emph{Flag codes} are a special family of multishot subspace codes, in which codewords are sequences of nested subspaces of a vector space over a finite field. In the network coding setting, they were first introduced in \cite{LiebNebeVaz18}. Given integers $1 \leq t_1 < t_2 < \dots < t_r < n$, a \emph{flag of type} $\type$ on $\bbF_q^n$ is a sequence of nested vector subspaces $\mathcal{F} = (\mathcal{F}_1, \dots, \mathcal{F}_r) \in \mathcal{G}_q(t_1, n) \times \dots \times \mathcal{G}_q(t_r, n)$ such that 
$$
\{0\} \subsetneq \mathcal{F}_1 \subsetneq \mathcal{F}_2 \subsetneq \dots \subsetneq \mathcal{F}_r \subsetneq \mathbb{F}_q^n.
$$
The $t_i$-dimensional subspace $\mathcal{F}_i$ is called the \emph{$i$-th subspace} of the flag $\cF$. When we consider flags of \emph{full type vector}, that is, $(1, \dots, n-1)$, we speak about \emph{full flags}. 

The \emph{flag variety} of type $\type$ on $\bbF_q^n$ is denoted by $\cF_q(\type, n)$ and it is the set of flags of the corresponding type. As a subset of $\cP_q(n)^r$, the flag variety can be seen as a metric space, equipped with the extended subspace distance given in (\ref{eq: extended subspace distance}). We call it the \emph{flag distance} and denote it by $d_f$ in this setting. More precisely, if $\cF=(\cF_1, \dots, \cF_r)$ and $\cF'=(\cF'_1, \dots, \cF'_r)$ are flags of type $\type$ on $\bbF_q^n$, the flag  distance between them is given by
$$
d_f(\cF, \cF') = \sum_{i=1}^r d_S(\cF_i, \cF'_i).
$$

A \emph{flag code} of type $\type$ on $\bbF_q^n$ is a nonempty subset $\cC$ of the flag variety  $\cF_q(\type, n)$ and its \emph{(minimum) distance} is given by
$$
d_f(\cC) = \min\{ d_f(\cF, \cF') \ | \ \cF, \cF'\in \cC, \ \cF\neq \cF \}.
$$
Observe that, if $\cC$ contains at least two flags, its distance is a positive even integer. On the other hand, if $|\cC|=1$, we put $d_f(\cC)=0$. The bounds for the distance between $t_i$-dimensional subspaces given in (\ref{eq:bound subspace codes}) yield to the following upper bound for the flag distance:
$$
d_f(\cC) \leq 2\left( \sum_{t_i\leq\lfloor\frac{n}{2}\rfloor} t_i + \sum_{t_i>\lfloor\frac{n}{2}\rfloor} (n-t_i) \right).
$$
Flag codes attaining this bound are called \emph{optimum distance flag codes} (see \cite{AloNavEsc19, AloNavEsc20}).

To finish this section, we present a channel for flags following the general idea of the channel model introduced in \cite{LiebNebeVaz18}. If we see flag codes as a particular case of multishot codes, sending a flag of type $\type$ on $\bbF_q^n$ (as a codeword of a flag code) requires using the subspace channel $r$ times to send $r$ nested subspaces of $\bbF_q^n$. The nested structure allows us to reduce the amount of sent information in every shot. Let us make this precise.

The network can be modelled as a finite directed acyclic multigraph with a single source and several receivers. Assume that we want to send a flag $\cF=(\cF_1, \dots, \cF_r)$. By virtue of the nested structure of flags, one can find vectors $v_1, \dots, v_{t_r} \in \bbF_q^n$ such that for every $1\leq i\leq r$, the subspace $\cF_i$ is spanned by the set of vectors
$$
\{v_1, \dots, v_{t_i}\}.
$$
In order to send the flag $\cF$, we proceed as follows. We fix $t_0=0$ and for every value of $1\leq i\leq r$, at the $i$-th shot:

\begin{itemize}
\item The source injects the set of vectors $\{v_{t_{i-1}+1}, \dots, v_{t_i}\}$. One of each through a different outgoing edge. 
\item Then, intermediate nodes construct random $\mathbb{F}_q$-linear combinations of the received vectors up to this moment (included the ones received in previous shots) and send each of them through an outgoing edge.
\item The receivers get random linear combinations of the vectors $\{v_1, \dots, v_{t_i}\}$ and put them as the rows of a matrix $Z_i$ to construct the subspace

$$
\cX_i=\mathrm{rowsp}
\begin{pmatrix}
Z_1\\
\vdots\\
Z_i
\end{pmatrix},
$$
that is, the vector space spanned by the rows of $Z_1, \dots, Z_r$.
\end{itemize}

After $r$ shots, every receiver is able to form a \emph{stuttering flag} $\cX= (\cX_1, \dots , \cX_{r})$, i.e., a sequence of nested subspaces where equalities are allowed. Note that, in absence of errors, this sequence coincides with the sent flag $\cF$. Nevertheless, \emph{erasures} and \emph{insertions} can occur at every shot. Let us explain in more detail these two concepts. If we write $\cX_i=\bar{\cF}_i\oplus \cE_i$,  with $\bar{\cF}_i\subseteq \cF_i$ and $\cF_i\cap\cE_i=\{0\}$, then the \emph{number of erasures at the $i$-th shot} is 
$$
d_S(\cF_i, \bar{\cF}_i)= \dim(\cF_i)-\dim(\bar{\cF}_i)
$$
and it represents the number of dimension losses from $\cF_i$ in $\cX_i$. The \emph{number of insertions at the $i$-th shot} is $\dim(\cE_i)$ and it measures the dimension of the vector space generated by the set of vectors in $\cX_i$ but not in $\cF_i$. Hence, the \emph{number of errors at the $i$-th shot} is given by $e_i=d_S(\cF_i, \cX_i)$ and it counts both the number of erasures and insertions occurred in this single shot. Finally, the \emph{total number of errors} of the communication is denoted by $e$ and it is computed as
$$
e=\sum_{i=1}^r e_i = \sum_{i=1}^r d_S(\cF_i, \cX_i) = d_f(\cF, \cX).
$$

In a general channel, this number $e$ tells us how many erasures and insertions have occurred in the transmission of the flag $\cF$. Through this paper, we will focus on a more particular channel, the \emph{erasure channel for flags}, in which insertions are not allowed. Hence, only erasures can occur and it holds $\cX_i \subseteq \cF_i$ for every $1\leq i\leq r$. In this context, the value $e$ is called \emph{total number of erasures} and $e_i$ is called the \emph{number of erasures at the $i$-th shot}.

\section{Consistent flag codes}\label{sec:Consistent flag codes}

Following the ideas in \cite{AloNavEsc19}, given a flag code $\cC,$ we can naturally associate to it a set of constant dimension codes that we obtain when we gather the subspaces of the same dimension used at a fixed shot in the process of sending flags. Let us explain more precisely the definition of these codes.

\begin{Definition}
Let $\cC$ be a flag code of type $\type$ on $\bbF_q^n$. For every $1\leq i\leq r$, we call the $i$-\emph{projected code} of $\cC$ to the constant dimension code $\cC_i \subseteq \cG_q(t_i, n)$ given by the set of $i$-th subspaces of flags in $\cC$. More precisely, 
$$
\cC_i = \{ \cF_i \ | \ (\cF_1, \dots, \cF_i, \dots, \cF_r) \in \cC \}.
$$
\end{Definition}

Due to the close relationship between a flag code $\cC$ and its projected codes, it is a natural question to explore how far properties and structure of these codes determine  the structure of $\cC$ and conversely. In this direction, this section is devoted to the study of flag codes that are \emph{consistent} with respect to their projected codes or just \emph{consistent flag codes}. This is a family of flag codes in which the cardinality and distance are completely determined by the ones of their projected codes. We will show how, in some specific cases, this property of consistency goes beyond size and distance and gives rise to a stronger structural consistency. Furthermore, as we will see in Section \ref{sec: Our decoding algorithm}, the property of being consistent makes it possible to give a decoding algorithm in the erasure channel for flags.

 Let us first point out the connection between the cardinality of a flag code $\cC$ of type $(t_1, \ldots, t_r)$ and the ones of its projected codes. It is clear that the size of any projected code is upper-bounded by the cardinality of $\cC$, that is,  
 $$|\cC_i|\leq |\cC|, \, i=1, \ldots, r.$$

The first condition of consistency we will impose to our family of flags is the property of \emph{disjointness}.

\begin{Definition}
A flag code $\cC$ of type $\type$ on $\bbF_q^n$ is said to be \emph{disjoint} if its cardinality coincides with the ones of its projected codes, that is, if 
$$
|\cC|=|\cC_1|=\dots=|\cC_r|.
$$
\end{Definition}
The notion of disjoint flag codes was introduced in \cite{AloNavEsc19} in order to characterize optimum distance flag codes in terms of their projected codes.
Observe that the cardinality of a disjoint flag code is  determined by the one of any of its projected codes. In this sense, we can say that disjoint flag codes are \emph{consistent} with respect to the cardinality of their projected codes, or just \emph{cardinality-consistent}, for short.

Just as the cardinality of a disjoint flag code is determined by the ones of its projected codes, we introduce the concept of consistency of flag codes with respect to the distance of their projected codes taking into account the pairs of flags attaining the minimum distance of the flag code. 

\begin{Definition}
Let $\cC$ be a flag code of type $\type$ on $\bbF_q^n$. We say that $\cC$ is \emph{distance-consistent} if for every pair of different flags $\cF, \cF'$ in $\cC,$ the following statements are equivalent:
\begin{enumerate}
\item $d_f(\cF, \cF')=d_f(\cC)$.
\item $d_S(\cF_i, \cF'_i)=d_S(\cC_i)$, for all $i=1, \dots, r.$
\end{enumerate}
\end{Definition}

Notice that, in a distance-consistent flag code, a pair of flags provides the minimum distance if, and only if, the (subspace) distance between their subspaces is the minimum (subspace) distance of the corresponding projected code. Hence, closest flags in a distance-consistent flag code are given by nested sequences of the closest subspaces in the projected codes. That does not occur   in a general flag code, as we can see in the following example: 

\begin{Example}\label{ex: non-consistent flag code}
Let $\cC$ be the flag code of type $(1,2,3)$ on $\bbF_q^5$ given by the set of flags
$$
\begin{array}{rcl}
\cF^1 & = & (\langle u_1 \rangle, \langle u_1, u_3 \rangle, \langle u_1, u_3, u_4 \rangle  ),\\  
\cF^2 & = & (\langle u_1 \rangle, \langle u_1, u_5 \rangle, \langle u_1, u_2, u_5 \rangle  ),\\  
\cF^3 & = & (\langle u_2 \rangle, \langle u_1, u_2 \rangle, \langle u_1, u_2, u_4 \rangle  ),
\end{array}
$$where $\{u_i\}_{i=1}^5$ denotes the standard basis of $\bbF_q^5$ over $\bbF_q$.

In this case, we have $d_f(\cC)=d_f(\cF^1, \cF^2)=6$. However, the flag code $\cC$ is not distance-consistent since the distance of every projected code is $d_S(\cC_i)=2,$ $i=1,2,3$, but $d_S(\cF_1^1, \cF_1^2)=0$ and $d_f(\cF_3^1, \cF_3^2)=4.$
\end{Example}

Under the distance-consistency condition, there is a coherent link between what is close both at flag level and at subspace level. Moreover, it follows a clear connection between the distance of a flag code and the ones of its projected codes.

\begin{Proposition}\label{prop: distance of a consistent flag code}
The distance of a distance-consistent flag code coincides with the sum of the ones of its projected codes.
\end{Proposition}

The previous example shows that the converse of this result is not true in general. Just observe that the distance of $\cC$ is $6,$ which is the sum of the distances of its projected codes, whereas $\cC$ is not distance-consistent. However, we can notice that, in this example, the minimum distance is attained by two different flags $\cF^1, \cF^2\in\cC$ with a common subspace. If we exclude this situation and focus on flag codes where different flags have all their subspaces different, i.e., disjoint flag codes, we have the following characterization.

\begin{Proposition}\label{prop: equivalencia distance-consistent}
Let $\cC$ be a disjoint flag code of type $\type$ on $\bbF_q^n$. The following statements are equivalent:
\begin{enumerate}
\item $\cC$ is distance-consistent;
\item $d_f(\cC)=\sum_{i=1}^r d_S(\cC_i).$
\end{enumerate}
\end{Proposition}
\begin{proof}
By means of Proposition \ref{prop: distance of a consistent flag code}, we just need to prove that a disjoint flag code $\cC$ with distance $d_f(\cC)=\sum_{i=1}^r d_S(\cC_i)$ must be distance-consistent. To do so, consider a pair of different flags $\cF, \cF'$ in $\cC$ giving the distance of the code. Since $\cC$ is disjoint, we know that $\cF_i\neq \cF'_i$ for every $1\leq i \leq r$. As a result, for every value of $i$, the distance $d_S(\cF_i, \cF'_i)$ cannot be zero and hence must be, at least, $d_S(\cC_i).$ On the other hand, we have that
$$
d_f(\cF, \cF')=d_f(\cC)=\sum_{i=1}^r d_S(\cC_i),
$$
which happens if, and only if, $d_S(\cF_i, \cF'_i)=d_S(\cC_i)$ for every $1\leq i\leq r$, that is, if $\cC$ is distance-consistent.
\end{proof}

These two concepts of consistency, with respect either to the cardinality or to the distance, give rise to a more general idea of consistency which gathers both of them. 

\begin{Definition}
A flag code is said to be \emph{consistent}  (w.r.t. its projected codes), if it is both cardinality-consistent (disjoint) and distance-consistent.
\end{Definition}

This definition along with Proposition \ref{prop: equivalencia distance-consistent} provides the following characterization of consistent flag codes. 

\begin{Theorem}\label{theo: caracterización consistente}
Let $\cC$ be a flag code of type $\type$ on $\bbF_q^n$. The following statements are equivalent:
\begin{enumerate}
\item The code $\cC$ is consistent.
\item The code $\cC$ is disjoint and $d_f(\cC)=\sum_{i=1}^r d_S(\cC_i).$
\end{enumerate}
\end{Theorem}

Observe that, by means of this result, in order to determine if a flag code is consistent, we just need to compute the distance and cardinality of the given flag code as well as the ones of its projected codes. This is notably easier than checking the distance-consistency condition, i.e., that every pair of flags in the code gives the minimum distance if, and only if, the distance between their subspaces coincides with the minimum distance of every projected code. In particular, every flag code consisting of a single flag is automatically consistent with $d_f(\cC)=0$. 

To finish this section, we deepen the structure of consistent flag codes in order to give some crucial definitions and properties for the design of the decoding algorithm described in Section \ref{sec: Our decoding algorithm}.  Let us fix $\cC$ a consistent flag code of type $\type$ on $\bbF_q^n$. Observe that, if $\cF, \cF'\in \cC$ attain the minimum distance of $\cC$, by virtue of the distance-consistency property, it holds 
$$
d_S(\cC_i)= d_S(\cF_i, \cF'_i)= 2(t_i - \dim(\cF_i\cap\cF'_i))
$$ for every $1\leq i \leq r.$ Equivalently, the distance of $\cC$ is attained by a pair of flags $\cF, \cF'\in \cC$ if, and only if, we have $\dim(\cF_i\cap\cF'_i)=m_i,$  where 
\begin{equation}\label{eq: fixed dim of the intersection} 
m_i = t_i - \frac{d_S(\cC_i)}{2},
\end{equation}
for every $1\leq i\leq r$.

Recall also that, as the code $\cC$ is consistent, in particular, it is disjoint. Hence, if $|\cC|\geq 2$, every projected code contains at least two different subspaces and its distance is a positive even integer. As a result, for every value $1\leq i\leq r$, we have that $m_i < t_i$. Moreover, since subspaces in a flag are nested, associated to $\cC$ we obtain a non-decreasing sequence of integers $0\leq m_1\leq \cdots \leq m_r < t_r$ such that,  for each pair $\cF, \cF'$ with $d_f(\cF, \cF')=d_f(\cC)$, we can construct a (stuttering) flag
$$\cF \cap \cF':=(\cF_1\cap \cF'_1, \dots, \cF_r\cap \cF'_r)$$
of type $(m_1, \dots, m_r)$. With this notation, we give the following definition.

\begin{Definition}
Let $\cC$ be a consistent flag code of type $\type$ on $\bbF_q^n$. We define the \emph{minimum distance intersection code} of $\cC$ as the stuttering flag code of type $(m_1, \dots, m_r)$ given by the family
\begin{equation}\label{eq: minimum distance intersection code}
\{ \cF\cap \cF' \ | \ d_f(\cF, \cF')= d_f(\cC) \}.
\end{equation}
\end{Definition}

Notice that, for not consistent flag codes, the set given in (\ref{eq: minimum distance intersection code}) has not necessarily a fixed type. For instance, if we consider the code $\cC$ in the Example \ref{ex: non-consistent flag code}, the set 
$$
\{ \cF\cap \cF' \ | \ d_f(\cF, \cF')= 6 \} = \{ \cF^1\cap \cF^2, \cF^1\cap \cF^3, \cF^1 \cap \cF^3 \}
$$
contains stuttering flags of types $(1, 1, 1)$ and $(0, 1, 2)$.

The sequence of numbers $(m_1, \dots, m_r)$ defined as in (\ref{eq: fixed dim of the intersection}) provides upper bounds for the dimension of the intersection of subspaces in every projected code of a consistent flag code.

\begin{Proposition}\label{prop: dim intersection disjoint+consistent}
Let $\cC$ be a consistent flag code $\cC$ of type $\type$ and consider a pair of different flags $\cF, \cF'\in \cC$. Then
$\dim(\cF_i \cap \cF'_i) \leq  m_i,$ for every $1\leq i \leq r$.
\end{Proposition}
\begin{proof}
Let $\cF$ and $\cF'$ be a pair of flags in a consistent flag code $\cC$. Since this code is disjoint, from the condition $\cF\neq \cF',$ we obtain that $\cF_i\neq \cF_i'$ for every value of $1\leq i\leq r$. Hence, it holds
$$
d_S(\cC_i) \leq d_S(\cF_i, \cF_i')= 2(t_i - \dim(\cF_i\cap\cF_i'))
$$
or, equivalently, 
$$
\dim(\cF_i\cap\cF_i') \leq t_i-\frac{d_S(\cC_i)}{2} = m_i.
$$
\end{proof}

Notice that, in the previous result, the condition of consistency cannot be relaxed in none of its two sides. On the one hand, being distance-consistent is necessary to properly define the numbers $m_1, \dots, m_r$. On the other hand, the next example shows that the condition of cardinality-consistency can neither be removed.
\begin{Example}
Consider the flag code $\cC$ of type $(1,2,3,4)$ on $\bbF_q^6$ consisting of the set of three flags
$$
\begin{array}{cccccc}
\mathcal{F}^1 &=& (\left\langle u_1 \right\rangle, & \left\langle u_1, u_2 \right\rangle , & \left\langle u_1, u_2, u_3 \right\rangle & \left\langle u_1, u_2, u_3, u_4 \right\rangle),\\
\mathcal{F}^2 &=& ( \left\langle u_2 \right\rangle, & \left\langle u_2, u_3 \right\rangle , & \left\langle u_2, u_3, u_4 \right\rangle & \left\langle u_2, u_3, u_4, u_5\right\rangle), \\
\mathcal{F}^3 &=& ( \left\langle u_1 \right\rangle , &  \left\langle u_1, u_5 \right\rangle, & \left\langle u_1, u_4, u_5 \right\rangle & \left\langle u_1, u_4, u_5, u_6 \right\rangle),
\end{array}
$$
where $\{u_1, \dots, u_6\}$ is the standard basis of $\bbF_q^6$ over $\bbF_q$.

Observe that $\cC$ is not disjoint since $\cC_1$ contains only two subspaces. The distance of every projected code is $2$ and the distance of the flag code is $d_f(\cC)=8$. This distance is only attained by the pair of flags $\cF^1$ and $\cF^2$ and, for this couple of flags, it holds $d_S(\cF_i^1, \cF_i^2)= 2 = d_S(\cC_i)$ for every $i=1, 2,3,4$. Hence, $\cC$ is a distance-consistent flag code and it makes sense to consider the values
$$
m_i= t_i - \frac{d_S(\cC_i)}{2} = i - 1, \ \text{for} \ 1\leq i\leq 4,
$$
defined as in (\ref{eq: fixed dim of the intersection}). However, notice that $\dim(\cF_1^1 \cap \cF_1^3)= 1 > 0 = m_1,$ in contrast to what happens in the context of Proposition \ref{prop: dim intersection disjoint+consistent}, where the extra condition of cardinality-consistency (disjointness) is required.
\end{Example}

\section{Some families of consistent flag codes}\label{sec: Some families of consistent flag codes}

Given that a flag code $\cC$ of type $(t_1, \ldots, t_r)$ can be seen as a subset of the product of the Grassmannians $\cG_q(t_1,n) \times \dots\times \cG_q(t_r,n),$ it seems quite natural, beyond cardinality and distance questions, to study the relationship between flag codes satisfying certain property and flag codes with projected codes fulfilling the equivalent property at the subspace codes level.  

With this objective in mind, in this section we introduce two couples of special families of flag codes. We start studying and comparing equidistant flag codes and flag codes with equidistant projected codes. We show that, in general, these concepts are not equivalent whereas under the assumption of being consistent they coincide. Next, we focus on the study of sunflower flag codes and flag codes with sunflower (subspace) codes as their projected codes. Again, we see that, in general, these two families do not coincide but they turn to be the same when we impose the consistency condition. This study allows us to conclude that the consistency condition defined in Section \ref{sec:Consistent flag codes} leads as well to a structural consistency that strongly relates the nature of a flag code with the ones of its projected codes. Let us make these ideas precise.

\subsection{Consistent equidistant flag codes}
Recall from Section \ref{sec:preliminares} that a subspace code $\cC$ in $\cG_q(k, n)$ is called equidistant if for each pair of codewords $\cU, \cV$ in $\cC$, we have that $d_S(\cU, \cV)=d_S(\cC)$. Let us generalize this concept to the flag codes scenario in two different ways.

\begin{Definition}
A flag code $\cC\subseteq \cF_q(\type, n)$ is said to be \emph{equidistant} if, for every pair of different flags $\cF, \cF'\in \cC$, it holds $d_f(\cC)=d_f(\cF, \cF')$.
\end{Definition}

\begin{Definition}
A flag code $\cC$ is said to be \emph{projected-equidistant} if all its projected codes $\cC_i$ are equidistant constant dimension codes. 
\end{Definition}

These two concepts do not represent the same notion in the general framework of flag codes as the reader can realize from the following two examples. First, we see that equidistant flag codes do not need to be projected-equidistant.
\begin{Example}
Let $\{u_1, ,\dots, u_5\}$ denote the standard basis of $\bbF_q^5$ over $\bbF_q$ and consider the flag code $\cC$ of type $(2,3)$ on $\bbF_q^5$, consisting of the set of flags
$$
\begin{array}{cccl}
\cF^1 & = & (\langle u_1, u_2\rangle, & \langle u_1, u_2, u_3\rangle),\\
\cF^2 & = & (\langle u_1, u_4\rangle, & \langle u_1, u_4, u_5\rangle),\\
\cF^3 & = & (\langle u_3, u_4\rangle, & \langle u_2, u_3, u_4\rangle).
\end{array}
$$
This code is equidistant with $d_f(\cC)=6$ but none of its projected code is equidistant. For instance, $d_S(\cF_1^1, \cF_1^2)=2\neq 4 = d_S(\cF_1^1, \cF_1^3)$.
\end{Example}

The other way round, being projected-equidistant does not imply equidistance.
\begin{Example}\label{ex: proj-equid is not equid}
Let $\cC$ be a full flag code on $\bbF_q^3$ given by the set of flags
$$
\begin{array}{cccl}
\cF^1 & = & (\langle u_1\rangle, & \langle u_1, u_3\rangle),\\
\cF^2 & = & (\langle u_1\rangle, & \langle u_1, u_2+u_3\rangle),\\
\cF^3 & = & (\langle u_2\rangle, & \langle u_1, u_2\rangle),
\end{array}
$$
where $\{u_1, u_2, u_3\}$ is the standard basis of $\bbF_q^3$ over the field $\bbF_q$. Observe that $\cC_1$ is and $\cC_2$ are equidistant constant dimension codes. In other words, the flag code $\cC$ is projected-equidistant. Nevertheless, it is clear that $\cC$ is not equidistant since 
$$
d_f(\cF^1, \cF^2)= 2 \neq 4 = d_f(\cF^1, \cF^3).
$$
\end{Example}

In light of Example \ref{ex: proj-equid is not equid}, we can see that projected-equidistant flag codes are not in general consistent since they might not even be cardinality-consistent (disjoint). When we require them to satisfy this extra condition, we obtain the following results.

\begin{Proposition}\label{prop: disjoint proj-equid are equid}
Let $\cC\subseteq \cF_q(\type, n)$ be a projected-equidistant flag code. If $\cC$ is cardinality-consistent, then it is equidistant with distance $d_f(\cC)=\sum_{i=1}^r d_S(\cC_i)$.
\end{Proposition}
\begin{proof}
Assume that $\cC$ is a cardinality-consistent projected-equidistant flag code, i.e., a disjoint flag code with equidistant projected codes. We distinguish two cases:
\begin{itemize}
\item If $|\cC|=1$, we have $|\cC_i|=1$ for each $1\leq i\leq r$. Hence, $d_f(\cC)=d_S(\cC_i)=0$ and the result trivially holds. 
\item If $|\cC|\geq 2$, consider an arbitrary pair of different flags $\cF, \cF'\in \cC$. The cardinality-consistency condition implies that $\cF_i\neq \cF'_i$ for every $1\leq i\leq r$. Now, given that every projected code $\cC_i$ is equidistant, we have that $d_S(\cF_i, \cF'_i)=d_S(\cC_i)$ for every $1\leq i\leq r$. Consequently, $d_f(\cC)=d_S(\cF, \cF')=\sum_{i=1}^r d_S(\cC_i)$ and the flag code $\cC$ is equidistant.
\end{itemize}
\end{proof}

As a consequence, and by means of Theorem \ref{theo: caracterización consistente}, we obtain the next corollary.
\begin{Corollary}\label{cor: disjoint proj-equid are consistent}
Disjoint projected-equidistant flag codes are consistent.
\end{Corollary}

\begin{Remark}
Observe that if $\cC\subseteq\cF_q(\type, n)$ is a projected-equidistant flag code, then there exist integers $c_1, \dots, c_r$ such that every projected code $\cC_i$ is an equidistant $c_i$-intersecting constant dimension code in $\cG_q(t_i, n)$. Moreover, if $\cC$ is disjoint, by means of the previous corollary, it is consistent. In this case, the sequence $(c_1, \dots, c_r)$ is precisely the type vector of the minimum distance intersection code defined as in (\ref{eq: minimum distance intersection code}) associated to the consistent flag code $\cC$.
\end{Remark}

According to these two results, it is clear that disjoint projected-equidistant flag codes are equidistant and consistent. In the next result, we prove that they are exactly the only codes that can be equidistant and consistent simultaneously.
\begin{Theorem}\label{theo: consistent + equidistant}
A consistent flag code is equidistant if, and only if, it is projected-equidistant.
\end{Theorem}
\begin{proof}
The ``if'' part is consequence of Proposition \ref{prop: disjoint proj-equid are equid}. For the ``only if'' part, assume that $\cC$ is a consistent and equidistant flag code. Let us see that, under these two conditions, every projected code is equidistant. In case $\cC$ contains a single flag, it is clear that every projected code is equidistant (with distance equal to zero).

Let us study the case $|\cC|\geq 2$. Observe that, since $\cC$ is consistent, in particular the cardinality of every projected code coincides with the one of $\cC$ and, given an index $1\leq i\leq r$, we can always find two different subspaces $\cU, \cV\in \cC_i$. By the definition of projected code, there must exist different flags $\cF, \cF'\in \cC$ such that $\cF_i=\cU$ and $\cF'_i=\cV$. Since $\mathcal{C}$ is equidistant, it holds $d_f(\cC)=d_f(\cF, \cF')$. This fact together with the distance-consistency property implies that $d_S(\cF_j, \cF'_j)=d_S(\cC_j)$ for every $1\leq j\leq r$. In particular, we conclude that $d_S(\cC_i)=d_S(\cF_i, \cF'_i)=d_S(\cU, \cV)$, which proves that the projected code $\cC_i$ is equidistant. 
\end{proof}

Observe that consistency completely translates the property of being equidistant from a flag code to its projected codes and vice versa. As a particular case of equidistant flag codes, we mention the family of optimum distance flag codes.

\subsubsection{Optimum distance flag codes}
Optimum distance flag codes were introduced in \cite{AloNavEsc19} as a generalization of constant dimension codes attaining the maximum distance. Constructions of them can be found as well in \cite{AloNavEsc20}. These codes were characterized as disjoint flag codes with projected codes attaining the maximum possible distance for the corresponding dimension. In particular, the projected codes of every optimum distance flag code are equidistant subspace codes. Hence, optimum distance flag codes are an example of disjoint projected-equidistant flag codes. As a result, by means of Corollary \ref{cor: disjoint proj-equid are consistent}, the next result holds.

\begin{Corollary}
Optimum distance flag codes are consistent.
\end{Corollary}

Once again, consistency allows to perfectly translate the property of attaining maximum distance at the subspace codes level to flag codes and conversely.

\subsection{Consistent sunflower flag codes}

Following the ideas in \cite{GoRav16}, here below we introduce the concept of \emph{sunflower} flag code.

\begin{Definition}
A flag code $\cC\subseteq\cF_q(\type, n)$ is said to be a \emph{sunflower} if there exists a stuttering flag $C=(C_1, \dots, C_r)$ such that, for every pair of different flags $\cF, \cF'\in\cC$ it holds 
$$
\cF\cap\cF'=(\cF_1\cap\cF'_1, \dots, \cF_r\cap\cF'_r)= (C_1, \dots, C_r)=C.
$$ 
In this case, the stuttering flag $C$ is called the \emph{center} of the sunflower flag code $\cC$.
\end{Definition}

By analogy with the concept of projected-equidistant flag code, we define now the family of flag codes having sunflowers as projected codes.

\begin{Definition}
A flag code $\cC\subseteq\cF_q(\type, n)$ is said to be \emph{projected-sunflower} if all its projected codes are sunflowers. In this situation, there exist subspaces $C_1, \dots, C_r$ such that every $i$-projected code $\cC_i$ of $\cC$ is a sunflower of center $C_i$. We say that $C_1, \dots, C_r$ are the \emph{centers} of the projected-sunflower flag code $\cC$. 
\end{Definition}

One has the following relationship between these two concepts.

\begin{Proposition}\label{prop: sunflower implies projected-sunflower}
Let $\cC\subseteq \cF_q(\type, n)$ be a sunflower flag code of center $C=(C_1, \dots, C_r)$. Then the code $\cC$ is a projected-sunflower flag code of centers $C_1, \dots, C_r.$
\end{Proposition}
\begin{proof}
Assume that $\cC$ is a sunflower flag code of type $\type$. If $|\cC|=1$, the result trivially holds. Suppose now that $|\cC|\geq 2$. For every index $1\leq i\leq r$, we must prove that the projected code $\cC_i$ is a sunflower of center $C_i$. We distinguish to possibilities.
\begin{itemize}
\item If $|\cC_i|=|\cC|\geq 2,$ we can find two different subspaces $\cU, \cV\in \cC_i$. Hence, there exist flags $\cF, \cF'\in \cC$ such that $\cF_i=\cU$ and $\cF'_i=\cV.$ Since $\cC$ is a sunflower of center $C$, we have
$$
\cU\cap \cV = \cF_i\cap\cF'_i = C_i,
$$
which proves that $\cC_i$ is a sunflower of center $C_i$.
\item In case of $|\cC_i|<|\cC|,$ there must exist two different flags $\cF, \cF'$ in $\cC$ such that $\cF_i=\cF'_i$. Since $\cC$ is a sunflower of center $C$, we have that $C_i=\cF_i\cap\cF'_i=\cF_i$. On the other hand, notice that every subspace in the projected code $\cC_i$ contains the subspace $C_i$. Using that $\dim(C_i)=t_i$, we conclude that  $\cC_i=\{C_i\},$ which is the trivial sunflower of center $C_i$. 
\end{itemize}
\end{proof}

The converse of Proposition \ref{prop: sunflower implies projected-sunflower} is not true in general. The next example shows that projected-sunflower flag codes are not necessarily sunflowers. 
\begin{Example}
Consider the flag code $\cC\subseteq \cF_q((2,3), 4)$ given by the set of flags
$$
\begin{array}{cccc}
\cF^1 & = & (\langle u_1, u_2 \rangle, & \langle u_1, u_2, u_3 \rangle ), \\
\cF^2 & = & (\langle u_1, u_3 \rangle, & \langle u_1, u_2, u_3 \rangle ), \\
\cF^3 & = & (\langle u_1, u_4 \rangle, & \langle u_1, u_2, u_4 \rangle ), \\
\end{array}
$$
where $\{u_1, u_2, u_3, u_4\}$ denotes the standard basis of $\bbF_q^4$ over $\bbF_q$.
Observe that the projected codes are sunflowers of center $C_1=\langle u_1 \rangle$ and $C_2=\langle u_1, u_2 \rangle.$ However, the code $\cC$ is not a projected sunflower of center $(C_1, C_2)$ since $\cF_2^1\cap\cF_2^2\neq C_2$.
\end{Example}

Notice that the code in the previous example is not disjoint (cardinality-consistent) since $\cF^1$ and $\cF^2$ have the same second subspace. If we require the cardinality-consistency condition, we have the following characterization.
\begin{Theorem}\label{theo: equivalencia sunflower}
Let $\cC\subseteq \cF_q(\type, n)$ be a cardinality-consistent flag code. The following statements are equivalent:
\begin{enumerate}
\item $\cC$ is a sunflower of center $(C_1, \dots, C_r);$ \label{item1}
\item $\cC$ is a projected-sunflower of centers $C_1, \dots, C_r$.  \label{item2}
\end{enumerate}
\end{Theorem}
\begin{proof}
By means of Proposition \ref{prop: sunflower implies projected-sunflower}, it suffices to see that (\ref{item2}) implies (\ref{item1}). Assume that $\cC$ is a projected-sunflower and, for every $1\leq i\leq r$,  denote by $C_i$ the center of $\cC_i$. Consider two different flags $\cF, \cF'\in \cC$. Since $\cC$ is disjoint, we get that $\cF_i\neq\cF'_i$ for every $1\leq i\leq r$. Hence, we have $\cF_i\cap\cF'_i=C_i$ for every choice of $i$ and we conclude that $\cC$ is a sunflower flag code of center $(C_1, \dots, C_r).$
\end{proof}

Observe that projected-sunflower flag codes are, in particular, projected-equidistant. Hence, by the previous theorem together with Corollary \ref{cor: disjoint proj-equid are consistent}, the next result holds straightforwardly.

\begin{Corollary}
Assume that $\cC$ is a cardinality-consistent sunflower flag code of center $C=(C_1, \dots, C_r)$. Then it is consistent and its minimum distance intersection code is given by its center $\{C\}$.
\end{Corollary}

\begin{Remark}
Notice that Theorem \ref{theo: consistent + equidistant} still holds true if we replace ``equidistant'' by ``sunflower''. However, in the latter case, the condition of being consistent can be relaxed to just the one of cardinality-consistency, as shown in Theorem \ref{theo: equivalencia sunflower}. This is due to the fact that distance-consistency is an inherent property of sunflower flag codes. More precisely, if $\cC\subseteq\cF_q(\type, n)$ is a sunflower of center $(C_1, \dots, C_r)$, then this center not only determines the distance of the flag code $\cC$ given by
$$d_f(\cC)= 2\sum_{i=1}^r (t_i-\dim(C_i)),$$
but also the one of every projected code $\cC_i,$ which is $d_S(\cC_i)=2(t_i-\dim(C_i)).$

Once again we see that, under the consistency condition, one can naturally translate properties between the flag codes and the subspace codes frameworks. In this case, we can identify the flag codes property of being a sunflower with the one of having sunflowers as projected codes.
\end{Remark}

\section{A decoding algorithm for consistent flag codes}\label{sec: Our decoding algorithm}

In this section we provide a decoding algorithm on the erasure channel for consistent flag codes. In particular, it can be applied to the families mentioned in the previous section. This algorithm generalizes the ideas given in \cite{AloNavEsc19}, where a decoding process for optimum distance flag codes was given. As we will see through this section, both distance-consistency and cardinality-consistency play a key role in this procedure and allow us to reduce the problem of decoding a flag code to the one of just decoding one of its projected codes. 

Let $\cC$ be a flag code of type $\type$ on $\bbF_q^n$. Assume we have sent a flag $\cF=(\cF_1, \dots, \cF_r)\in \cC$ through the erasure channel for flags defined in Section \ref{subsec: flags and channel} (see \cite{LiebNebeVaz18}) and, after $r$ shots, a receiver gets a stuttering flag 
$$
\cX=(\cX_1, \dots, \cX_r). 
$$
As we are using an erasure channel, only erasures (no insertions) can occur. Hence, every subspace $\cX_i$ is contained in the subspace $\cF_i$ sent at the corresponding shot. Recall  from Subsection \ref{subsec: flags and channel} that, at every shot, some information can be lost. This amount of information is called  {\em number of erasures at the $i$-th shot} and computed as $e_i=d_S(\cF_i, \cX_i)=\dim(\cF_i)- \dim(\cX_i)$. The {\em total number of erasures} of the communication is given by
$$
e= \sum_{i=1}^r e_i = d_f(\cF,\cX)
$$
and it is said to be \emph{correctable} by the flag code $\cC$ if
$$
e\leq \left\lfloor \frac{d_f(\cC)-1}{2} \right\rfloor.
$$
In this case, the received sequence $\cX$ can be decoded into $\cF$ by minimum distance in $\cC$. This means that $\cF$ is the closest flag in $\cC$ to the received stuttering flag $\cX$. Analogously, for every $1\leq i\leq r$, the number of erasures at the $i$-th shot $e_i$ is said to be \emph{correctable} by the projected code $\cC_i$ of $\cC$ if
$$
e_i\leq \left\lfloor \frac{d_S(\cC_i)-1}{2} \right\rfloor.
$$

The next proposition relates the correctability of the total number of erasures by a flag code $\cC$ and with the number of erasures occurred at each shot, under certain conditions on the distance of the code.

\begin{Proposition}\label{AlMenosUnoDecodificaBien}
Let $\cC$ be a flag code of type $\type$ on $\bbF_q^n$ and suppose that $d_f(\cC)\leq \sum_{i=1}^r d_S(\cC_i)$. If the total number of erasures $e$ is correctable by $\cC$, then there is some $e_i$ that is correctable by the corresponding $\cC_i.$ 
\end{Proposition} 
\begin{proof}
Assume that $e\leq \left\lfloor \frac{d_f(\cC)-1}{2}\right\rfloor$ but no $e_i$ is correctable, that  is,  $e_i > \frac{d_S(\mathcal{C}_i)}{2}-1,$ for all $1\leq i\leq r$. Thus, $e_i \geq \frac{d_S(\cC_i)}{2}$ and the total number of erasures satisfies
$$
e=\sum_{i=1}^{r} e_i \geq \sum_{i=1}^{r} \frac{d_S(\mathcal{C}_i)}{2} \geq \frac{d_f(\mathcal{C})}{2} > \left\lfloor \frac{d_f(\cC)-1}{2} \right\rfloor,
$$ which is a contradiction, since $e$ is correctable.
\end{proof}

The next result provides a characterization of a correctable number of erasures at some shot in terms of the dimension of the received subspace. The proof follows straightforwardly from the definition of the number of erasures at some shot, together with the condition of correctability. 

\begin{Proposition}\label{prop: correctability in terms of dimension}
The number of erasures at the $i$-th shot $e_i$ is correctable if, and only if, the dimension of the subspace $\cX_i$ is greater than $t_i-\frac{d_S(\cC_i)}{2}$.
\end{Proposition}

Observe that distance-consistent flag codes introduced in Section \ref{sec:Consistent flag codes}, by means of Proposition \ref{prop: distance of a consistent flag code}, satisfy the required condition on the distance of Proposition \ref{AlMenosUnoDecodificaBien}. Moreover, the quantity $t_i-\frac{d_S(\cC_i)}{2}$ that appears in Proposition \ref{prop: correctability in terms of dimension} is precisely the value $m_i$ defined in (\ref{eq: fixed dim of the intersection}). Hence, assuming that a correctable number of erasures have occurred, we can always decode by minimum distance at least one subspace $\cF_i$ of the sent flag $\cF$ in a distance-consistent flag code. If we add the condition of cardinality-consistency, we obtain the next result.

\begin{Theorem}
Let $\cC$ be a consistent flag code of type $\type$ on $\bbF_q^n$ and assume that  $e$, the total number of erasures of the communication, is correctable. Then there exists some $1\leq j \leq r$ such that $e_j$ is correctable and we can recover the sent flag $\cF$ as the unique flag in $\cC$ such that $\cX_j$ is contained in its $j$-th subspace.
\end{Theorem}
\begin{proof}
Observe that, by the property of distance-consistency, the distance of the code is $d_f(\cC)=\sum_{i=1}^r d_S(\cC_i)$. Hence, since $e$ is correctable, by applying Proposition \ref{AlMenosUnoDecodificaBien}, the number of erasures at some shot, say the $j$-th one, must be correctable as well. Thus, from Proposition \ref{prop: correctability in terms of dimension}, it holds
$$
\dim(\cX_j)> t_j - \frac{d_S(\cC_j)}{2} = m_j.
$$
Recall that, since we are sending flags through an erasure channel, the subspace $\cX_j$ is contained in $\cF_j$, i.e., the $j$-th subspace of the sent flag. Moreover, by means of Proposition \ref{prop: dim intersection disjoint+consistent}, $\cF_j$ is the unique subspace in $\cC_j$ containing $\cX_j$. Finally, given that $\cC$ is a disjoint flag code, we can recover $\cF$ as the unique flag in $\cC$ having $\cF_j$ as its $j$-th subspace.
\end{proof}

Observe that this result guarantees the success of this decoding algorithm, starting from an index $j$ such that the number of erasures $e_j$ is correctable. Such an index can be easily identified just by checking the dimensions of the received subspaces $\cX_1, \dots, \cX_r$ and applying Proposition \ref{prop: correctability in terms of dimension}. Observe that we do not need to wait to receive the whole stuttering flag $\cX$ to start to decode. Our idea is doing it sequentially during the transmission process. At every shot, we check the dimension of the received subspace until we obtain a subspace $\cX_j$ of dimension greater than $m_j=t_j-\frac{d_S(\cC_j)}{2}$. At that moment, we can easily recover $\cF_j$ and determine the flag $\cF$ in $j\leq r$ shots. We sum up these ideas in the following algorithm.

\vspace{-10pt}
\begin{quote}

\hrulefill \\
\vspace{-15pt}
\begin{center}
\textsc{Decoding algorithm}
\end{center}
\vspace{-15pt}
\hrulefill

\vspace{-5pt}

\begin{small}

{\textsc{Assumptions:} We send a flag $\cF$ in a consistent flag code $\cC$ of type $\type$ on $\bbF_q^n$. At each shot, the $i$-th subspace $\cF_i$ is sent and a subspace $\cX_i$ is received. The total number of erasures $e$ is correctable.}
\vspace{-10pt}

\hrulefill

\vspace{-5pt}

\noindent\textsc{Input:}  The received stuttering flag $\cX=(\cX_1, \dots, \cX_{r}).$\\
\noindent\textsc{Output:} The sent flag $\cF=(\cF_1, \dots, \cF_r) \in \cC.$

\vspace{-10pt}

\hrulefill

\vspace{-5pt}

\begin{itemize}
\item[] Define $i=1:$
\item[] if $\dim(\cX_i) > m_i = t_i - \frac{d_S(\cC_i)}{2}$,
\begin{itemize}
\item[] \textbf{then} decode $\cX_i$ into the only $\cF_i \in \cC_i$ that contains $\cX_i.$
\item[] \textbf{return:} the unique flag $\cF \in \cC$ that has $\cF_i$ as its $i$-th subspace.
\end{itemize}
\item[] else $i:= i+1.$
\end{itemize}
\vspace{-18pt}
\end{small}
\hrulefill
\end{quote}

This decoding process takes advantage of the consistency condition in two ways. First, under the assumption of a correctable number of erasures, the distance-consistency property makes it possible to reduce the problem of decoding $\cX$ into $\cF$ to the one of decoding some $\cX_i$ into the corresponding sent subspace $\cF_i$. After that, the cardinality-consistency condition allows us to come back to the flags setting and recover $\cF$ from any of its subspaces. Again, the use of the consistency property in flag codes transfers a problem at the flag codes level -the one of decoding on the erasure channel- to the equivalent problem in the subspace codes scenario.

\section{Conclusions and future work}

In this paper we have introduced the concept of consistency for flag codes. This new notion allows us to measure in some way how far the parameters and properties of a flag code are determined by the ones of its projected codes. Moreover, in our search for families of consistent flag codes, we have generalized  the concepts of equidistant and sunflower code to the flag codes framework in several ways and proved that, under the assumption of consistency, they coincide. In this way, consistency plays an important role in the study of properties of flag codes that can be induced by their projected codes and vice versa. A decoding algorithm for consistent flag codes in the erasure channel has been provided. 

In future works, we would like to generalize the decoding algorithm presented in this paper in two possible ways. On the one hand, exploring how to decode consistent flag codes in a general channel where insertions were allowed. On the other hand, we want to study the decoding process for more general families of flag codes on the erasure channel, by relaxing the property of consistency in one of its two sides: either the distance-consistency or the cardinality-consistency.

\end{document}